\documentclass[english,11pt,aps,prd,a4paper,preprintnumbers, floatfix,nofootinbib,showpacs,superscriptaddress,  notitlepage]{revtex4-1} 
 \pdfoutput=1
\usepackage[usenames,dvipsnames]{color}  %%%%%%% usenames,dvipsnames required to get BrickRed
\usepackage{graphicx}
\usepackage{setspace}
\usepackage{upgreek}
\usepackage{braket}
\usepackage{enumitem}
\usepackage{caption}
\usepackage{stackengine}
\usepackage{subcaption}
\usepackage{amssymb}
\usepackage{amsmath}
\usepackage{amssymb}
\usepackage[colorlinks=true,citecolor=darkred,urlcolor=darkred, pdfborder={0 0 0}]{hyperref}
\usepackage[normalem]{ulem}
\usepackage[table]{xcolor}
\usepackage{graphicx}
\makeatletter
\def\p@subsection{}
\makeatother
\usepackage{float}
\usepackage{orcidlink}

\usepackage{ctable}
\usepackage{colortbl}
\definecolor{darktangerine}{rgb}{1.0, 0.66, 0.07}
\definecolor{gainsboro}{rgb}{0.86, 0.86, 0.86}
\definecolor{ceruleanblue}{rgb}{0.16, 0.32, 0.75}
\usepackage{fdsymbol}
\usepackage[capitalize]{cleveref}
\usepackage{soul}

\usepackage{colortbl}
\usepackage{xcolor}
\usepackage{xfrac}

\usepackage{booktabs}
\definecolor{darkred}{rgb}{0.6,0,0}

\definecolor{linkcolor}{rgb}{0,0,0.5}

\definecolor{vdrgreen}{rgb}{0.0, 0.7, 0.0}

\def\gsim{\raise0.3ex\hbox{$\;>$\kern-0.75em\raise-1.1ex\hbox{$\sim\;$}}}
\def\lsim{\raise0.3ex\hbox{$\;<$\kern-0.75em\raise-1.1ex\hbox{$\sim\;$}}}

\def\beqn#1{\begin{equation}\label{#1}}
\def\eeqn{\end{equation}}

\def\beqa#1{\begin{eqnarray}\label{#1}}
\def\eeqa{\end{eqnarray}}

\usepackage{caption}
\usepackage{subcaption}
\usepackage{adjustbox}
\def\Z2{$\mathcal{Z_2}$}

\usepackage{bbold}
\usepackage{ragged2e}
\newcommand {\ignore}[1]{}

\usepackage{mathrsfs}
\makeatletter
\newcommand\footnoteref[1]{\protected@xdef\@thefnmark{\ref{#1}}\@footnotemark}
\makeatother
\def\321{$\mathrm{SU(3) \otimes SU(2) \otimes U(1)}$ }

\newcommand{\AddrCinvestav}{%
Departamento de F\'{\i}sica, Centro de Investigaci\'on
  y de Estudios Avanzados del IPN,\\ Apartado Postal 14-740 07000 Ciudad de Mexico, Mexico}

\usepackage{appendix}
\begin{document}

\title{\textcolor{black}{ Testing coherent elastic neutrino-nucleus scattering $N$ dependence }}

\author{Laura Duque~\orcidlink{0009-0009-6109-7946}}\email{laura.duque@cinvestav.mx}
\affiliation{\AddrCinvestav}

\author{O. G. Miranda~\orcidlink{0000-0002-0310-7060}}\email{omar.miranda@cinvestav.mx}
\affiliation{\AddrCinvestav}

\author{G. Sanchez Garcia~\orcidlink{0000-0003-1830-2325}}%
\email{g.sanchez@ciencias.unam.mx}%
\affiliation{Departamento de F\'{i}sica, Facultad de Ciencias,  Universidad Nacional Aut\'onoma de M\'exico,
Apartado Postal 70-542, Ciudad de M\'exico 04510, M\'exico}%

\begin{abstract}
Despite its elusive character, coherent elastic neutrino-nucleus scattering (CEvNS) detection is evolving as a precision measurement from different neutrino measurements. In this note we show how CEvNS experiments using the Spallation Neutron Source (SNS) are robust measurements that enter into a precision era. Despite the interplay of nuclear and elementary particle physics, we demonstrate that CEvNS current measurements are accurately described by using only one extra parameter, the number of neutrons, $N$. By making this test, we illustrate that current CEvNS are precise enough to reconstruct the number of neutrons in the nuclear target. At the same time, determining the number of nucleons in the target material can be used as a benchmark for the precision and accuracy of this kind of detectors.
\end{abstract}

\maketitle
%\newpage
%\tableofcontents

%%%%%%%%%%%%%%%%%%%%%%%%%%%%%%%%%
%%%%%%%%%%%%%%%%%%%%%%%%%%%%%%%%%
%%%%%%%%%%%%%%%%%%%%%%%%%%%%%%%%%
\section{Introduction}
%%%%%%%%%%%%%%%%%%%%%%%%%%%%%%%%%
%%%%%%%%%%%%%%%%%%%%%%%%%%%%%%%%%
%%%%%%%%%%%%%%%%%%%%%%%%%%%%%%%%

The new era of neutrino  physics precision is seeing the flourish of experimental measurements in different neutrino channels. It is clear that neutrino oscillation experiments are at the forefront of this new development~\cite{JUNO:2025gmd, JUNO:2026tap,  DUNE:2020lwj, DUNE:2020lwj, Hyper-Kamiokande:2018ofw}, but they are not the only ones  experiencing such a high development; for instance, the IceCube neutrino observatory~\cite{IceCube:2013low} have probed that neutrinos can also be a useful tool to observe different processes that may not be reachable with other techniques. 
Additionally, the recent rise of Coherent Elastic Neutrino-Nucleus Scattering (CEvNS) is an example of the variety of interesting reactions in the field, providing rich phenomenological analyses regarding probes of the SM~\cite{AtzoriCorona:2025xwr, Alpizar-Venegas:2025wor, AtzoriCorona:2024rtv, AtzoriCorona:2026wbu, AristizabalSierra:2026rlo, Maity:2024aji}, nuclear physics~\cite{Cadeddu:2017etk, Sierra:2023pnf, AtzoriCorona:2023ktl, Rossi:2023brv}, and constraints to new physics in the neutrino sector~\cite{DeRomeri:2022twg, DeRomeri:2026prc, AtzoriCorona:2026wbu, CONUS:2026uhz, Chattaraj:2025fvx, Breso-Pla:2025pds, Lozano:2025ekx, TEXONO:2025sub, Demirci:2025qdp, Liao:2025hcs, Li:2024iij}. CEvNS was first envisaged in the early 70's~\cite{Freedman:1973yd}, more as a conceptual idea within the Standard Model (SM) than as a process to be measured. However, after the development of low energy detector technologies, its measurement appeared to be feasible and became a reality after the first detection by the COHERENT collaboration in 2017 using a CsI detector exposed to the neutrino flux from the Spallation Neutron Source (SNS) at the Oak Ridge National Laboratory~\cite{COHERENT:2017ipa}. Almost ten years later, several other measurements have been reported by the same collaboration~\cite{COHERENT:2021xmm, COHERENT:2020iec, COHERENT:2026yje}, the most recent using a germanium detector~\cite{COHERENT:2026yje, COHERENT:2024axu}. Moreover, experiments using reactor neutrino facilities~\cite{Colaresi:2022obx, Ackermann:2025obx} and solar neutrinos from the $^8$B line~\cite{PandaX:2024muv, XENON:2024ijk, LZ:2025igz, XENON:2026ydt} have also reported evidence of CEvNS from these sources. 

The plethora of current experiments listed before and future prospects in spalattion sources~\cite{Baxter:2019mcx, CCM:2021leg, Collar:2025sle, Su:2023klh} and reactor facilities~\cite{CONNIE:2019swq, Wong:2005vg, MINER:2016igy, NUCLEUS:2019igx, Billard:2016giu} measuring CEvNS point towards a precision era for this interaction. Hence, different tests regarding the accuracy and agreement between experimental data and theoretical prediction is needed. Indeed, several analyses proving SM parameters like the weak mixing angle have been widely reported in the literature~\cite{DeRomeri:2026dac, AtzoriCorona:2025xwr, Alpizar-Venegas:2025wor}, showing within error bars a good agreement with the SM prediction at low energies. In addition, CEvNS interactions have been used to extract nuclear parameters such as the neutron rms radius of the scattering materials~\cite{Cadeddu:2017etk, Sierra:2023pnf, AtzoriCorona:2023ktl, Rossi:2023brv}, providing a powerful tool to determine this parameter through weak interactions. 

Despite being an interaction that must take into account both particle physics and nuclear phenomena, it is well known that the main characteristic signal of CEvNS comes from its effective dependence on the squared number of neutrons of the target material. This provides an alternative test for this interaction, by correctly reproducing the number of neutrons of the target material from experimental data. Hence, in this note we propose to analyze the accuracy of CEvNS experimental measurements by testing the $N$ dependence of its associated cross section. To this end, we replace the number of neutrons within the cross section by an effective parameter which value will be determined by  comparing the expected number of events against experimental data. In this way, the accuracy of the interaction will be apparent by correctly reproducing the corresponding number of neutrons for each target material. Unlike the weak mixing angle, this approach allows us to test CEvNS with a quantity that is independent of the energy regime of the interaction, and, in contrast to the neutron rms radius, we are testing a parameter for which we know the result in advance.

For the purpose of this work, we restrict ourselves to the case of measurements from SNS sources. However, this analysis can be extended to the case of solar neutrinos once statistics are competitive to SNS experiments. For reactor neutrinos, current experiments use Ge as a target material, with quenching factor shape and uncertainties at low energies having a significant impact on the predicted number of events. Hence, we only consider the Ge detector at the SNS. The remainder of this letter is as follows: In Section 2 we provide the theoretical framework for CEvNS tests, in Section 3 we precent our analysis procedure, providing the corresponding results in Section 4, and finally we give our conclusions in Section 5.

\section{Theoretical framework \label{theory}}
\subsection{CEvNS cross section}
To show the level of accuracy currently achieved by CEvNS measurements, we start by showing the dependence of the corresponding cross section on the number of neutrons. 
The CEvNS differential cross section within the SM is given by
\begin{equation}
    \frac{d\sigma}{dE_{nr}}\Bigg|_{SM}=\frac{G_F^2\,M}{\pi}\,F_W^2(|\vec{q}|^2)\,\left(Q_V^{\textrm{SM}}\right)^2\left(1- \frac{ME_{nr}}{2E_\nu^2}\right),
    \label{eq:cross}
\end{equation}
where $G_F$ is the Fermi coupling constant, $M$ is the mass of the target nucleus, $E_{nr}$ is the nuclear recoil energy, $E_\nu$ is the incident neutrino energy, and $Q_V^{\textrm{SM}}$ is the SM weak nuclear charge,
\begin{equation}
    Q_V^{\textrm{SM}}=g_V^p\,Z\,+g_V^n\,N
    \label{eq:charge}
\end{equation}
with $Z$ and $N$ the number of protons and neutrons within the nucleus, respectively. The terms $g_V^p=\frac{1}{2}-2\,\sin^2\theta_W$ and $g_V^n=-\frac{1}{2}$ are the vector coupling constants for  protons and neutrons, with $\theta_W$ the weak mixing angle. The nuclear form factor, $F_W$, is introduced in Eq.~\eqref{eq:cross} to take into account the spatial distribution of nucleons and the  loss of coherence at high momentum transfers. In this work we consider the Klein-Nystrand parameterisation~\cite{Klein:1999qj}:
\begin{equation}
    F_W(|\vec{q}|^2)=3\,\frac{j_1(|\vec{q}|R_A)}{|\vec{q}|R_A}\,\left[1+\left(\vec{q}\,a_k\right)^2\right]^{-1},
\end{equation}
where $j_1$ denotes the spherical Bessel function of first kind, $R_A$ is the effective nuclear radius of the target nucleus, parameterized as $R_A\approx1.2\,A^{1/3}$ fm, and $a_k = 0.7$ fm is the nuclear skin thickness.
\subsection{N-dependence in CEvNS}

From the theoretical treatment presented on the previous section, we see that the neutron number, $N$, has a strong presence in the CEvNS cross-section that mainly comes from the SM weak nuclear charge, $Q_V^{\textrm{SM}}$, and from the nuclear form factor. Because the proton's vector coupling is heavily suppressed by the weak mixing angle, we have that $g_V^p \ll |g_V^n|$, and the weak charge in Eq.~\eqref{eq:charge} is almost entirely determined by the number of neutrons in the scattering nucleus. As a result, the overall scattering rate effectively scales as $N^2$, which is the representative signature of this process. Our main goal is to show how the current experiments confirm this rule and, at the same time, use this rule as a benchmark showing the accuracy and precision that CEvNS experiments at spallation sources have reached. For this purpose, we replace $N$ in Eq.~\eqref{eq:cross} by an effective parameter, $\mathfrak{n}$, that will be confronted to experimental data. Physically, this parameter will be interpreted as the extracted weighted average number of neutrons in a given detector.

Notice that the $N$-dependence goes beyond the weak nuclear charge in the cross section as it is also present in both the nuclear form factor and kinematic terms. As seen in Eq.~\eqref{eq:cross}, the cross section is proportional to the nuclear mass, $M$, which basically scales linearly with the total number of nucleons, $A = Z + N$. To write the expected number of events in terms of our introduced effective parameter, we will consider the average mass of the nucleus to be $M = Z m_p + \mathfrak{n} m_n$, with $m_p$ and $m_n$ the mass of the proton and neutron, respectively. Notice that in our effective analysis, $\mathfrak{n}$ is not constrained to be an integer number and, while a large number of neutrons increases the probability of interaction according to the $N^2$ rule, the corresponding large nuclear mass constrains the resulting nuclear recoil energy, pushing the recoil spectrum towards low energies. Furthermore, nuclear effects are also $N$-dependent. Considering, for instance, the Klein-Nystrand form factor, we will have a dependence on the effective nuclear radius $R_A\approx1.2\,A^{1/3} = 1.2 (Z + \mathfrak{n})^{1/3}$. Therefore, for highly neutron-rich nuclei, this larger physical size of the target nucleus causes a more rapid loss of coherence as the transferred momentum increases. Consequently, while increasing $N$ amplifies the CEvNS cross section, it simultaneously dictates the kinematics and spatial constraints that suppress the number of events at high recoil energies.

\section{Analysis \label{sec:analysis}}
In this section we present the statistical analysis used to test the $N^2$ dependence of the CEvNS cross section. We focus on the three different target nuclei used by the COHERENT collaboration, namely CsI~\cite{COHERENT:2021xmm}, LAr~\cite{COHERENT:2020iec}, and Ge~\cite{COHERENT:2026yje}. For each case, we compute the expected number of signal events on the $i$th energy bin using
\begin{equation}\label{eq:Nevents_CEvNS}
N_{i,l}^{\textrm{S}}
=
\mathcal{E}
\int_{E_{\mathrm{nr}}^{i}}^{E_{\mathrm{nr}}^{i+1}}
\hspace{-0.3cm}
d E_{\mathrm{nr}}\,
\epsilon_E(E_{\mathrm{nr}})
\int_{0}^{E^{\prime\text{max}}_{\text{nr}}}
\hspace{-0.3cm}
dE'_{\text{nr}}
\,
P(E_{\text{nr}},E'_{\text{nr}})
 \int_{E_\nu^{\text{min}}(E'_{\text{nr}})}^{E_\nu^{\text{max}}}
\hspace{-0.3cm}
d E_\nu
\frac{d \phi_{l}}{d E_\nu}(E_\nu)
\frac{d\sigma} {dE'_\mathrm{nr}}\Big|_\mathrm{SM}(E_\nu, E'_{\mathrm{nr}})
,
\end{equation}
with $\mathcal{E} = N_{\mathrm{A}} m_{\mathrm{det}} / M$ the number of target nuclei on the detector of mass, $m_{\textrm{det}}$, and $N_A$ the Avogadro's number. We distinguish between the reconstructed and the real energies, denoted as $E_{\textrm{nr}}$ and $E'_{\textrm{nr}}$, respectively, with   $\epsilon_E(E_{\textrm{nr}})$ the efficiency of the detector, and $P(E_{\textrm{nr}}, E'_{\textrm{nr}})$ the smearing. Details on each of these quantities will be discussed below. The the neutrino flux, $d\phi_l/dE_\nu$, is given by the standard contributions from the decay of pions and muons at rest~\cite{DeRomeri:2022twg}, with $l = \nu_\mu, \nu_e, \bar\nu_\mu$. These fluxes are normalized to $\eta = rN_\textrm{POT}/4\pi L^2$, with $r$ the neutrinos per flavour, $N_{\textrm{POT}}$ the protons on target, and $L$ the distance from source to detector. These quantities are summarized in Tab.~\ref{tab:detectors} for the three different experimental setups.   

\begin{table}[b]
\centering
\begin{tabular}{|c|c|c|c|c|c|}
\hline
\textbf{~~Detector~~} & ~~~$m_{det}$ (kg)~~~ & ~~~$L$ (m)~~~ & ~~~~~$r$~~~~~ & ~~~$N_{POT}$ ($\times 10^{23}$)~~~ & ~~~Reference~~~ \\ \hline
\textbf{CsI} & 14.6 & 19.3 & 0.0848 & 3.198 & \cite{COHERENT:2021xmm} \\ \hline
\textbf{LAr} & 24 & 27.5 & 0.09 & 1.38 & \cite{COHERENT:2020iec} \\ \hline
\textbf{Ge} & 19.2 & 8.53 & 0.37 & 0.468 & \cite{COHERENT:2026yje} \\ \hline
\end{tabular}
\caption{Parameters used to compute the expected number of events on the three COHERENT experiments.}
\label{tab:detectors}
\end{table}

\subsection{Cesium Iodide}

We begin our analysis with the CsI data reported in Ref.~\cite{COHERENT:2021xmm}. Details on the energy efficiency and smearing can be found on the supplemental material of the same reference. For this case the collaboration released the distribution of events in both energy and reconstructed arrival time, $t_{\textrm{rec}}$, bins~\cite{COHERENT:2021xmm}. Hence, we compute the predicted number of events in a bi-dimensional grid given by 
\begin{equation}
\label{eq:Nevents_CEvNS_ij}
 N_{ij}^{\textrm{S}} = \sum_{l} \int_{t_{\mathrm{rec}}^{j}}^{t_{\mathrm{rec}}^{j+1}} dt_{\mathrm{rec}} f^n_T(t_{\mathrm{rec}}, \alpha_3)\epsilon_T(t_{\mathrm{rec}}) N_{i,  l}^{\textrm{S}},
\end{equation}
with $\epsilon_T(t_{\textrm{rec}})$ the time efficiency, also found in the supplemental material of Ref.~\cite{COHERENT:2021xmm}, and $f_T(t_{\textrm{rec}})$ the timing distribution for neutrinos produced in pion and meson decays at rest~\cite{Picciau:2022xzi}\footnote{The dependence on the parameter $\alpha_3$ will be explained below.}. After computing the expected CEvNS signal, $N_{ij}^{\textrm{S}}$, in energy and time bins, we test the $N^2$ dependence on the cross section by comparing the predicted events with the experimental data using the Poissonian function~\cite{DeRomeri:2022twg}
\begin{equation}
  \chi^2 = 2\sum_{i,j}\left [ N^{\textrm{th}}_{ij}(\mathfrak{n}) - N^{\textrm{exp}}_{ij} + N^{\textrm{exp}}_{ij}\ln\left ( \frac{N^{\textrm{exp}}_{ij}}{N^{\textrm{th}}_{ij}(\mathfrak{n})} \right )\right ] + \sum_{m=1}^2 \frac{\alpha_{m}}{\sigma_{\alpha_m}^2} + \sum_{k=1}^3\frac{\beta_k}{\sigma_{\beta_k}^2} ,
    \label{eq:chi:CsI}
\end{equation}
where $N_{ij}^{\textrm{exp}}$ corresponds to the experimental data (see appendix A of Ref.~\cite{DeRomeri:2022twg}), and $N_{ij}^{\textrm{th}}$ is the predicted number of events considering background contributions from steady-sate background (SSB), beam related neutrons (BRN), and neutrino induced neutrons (NIN)~\cite{COHERENT:2017ipa}, that is,
\begin{equation}
    N^{\textrm{th}}_{ij}(\mathfrak{n}) = (1+\alpha_1) N^{S}_{ij}(\mathfrak{n},\alpha_2,\alpha_3) +  (1+\beta_1)N^{\textrm{SSB}}_{ij}+  (1+\beta_2)N^{\textrm{BRN}}_{ij}(\alpha_3)+  (1+\beta_3)N^{\textrm{NIN}}_{ij}(\alpha_3) \,.
    \label{N:chi:CsI}
\end{equation}
Details on the computation of backgrounds are given in Ref.~\cite{DeRomeri:2022twg}. Notice that we have explicitly shown the dependence on the parameter, $\mathfrak{n}$, which we will use to prove the $N^2$ dependence on the cross section. The $\alpha_m$ in Eqs.~\eqref{eq:Nevents_CEvNS_ij} and \eqref{N:chi:CsI} are nuisance parameters with respect to which we minimize Eq.~\eqref{eq:chi:CsI}, with $\alpha_1$ accounting for flux normalization and quenching factor uncertainties, $\alpha_2$ for efficiency, and $\alpha_3$ for the timing uncertainty, see Ref.~\cite{DeRomeri:2022twg} for details. On the other hand, $\beta_k$ are nuisance parameters related to purely background normalization uncertainties, where we have used $k = 1, 2, 3$ for SSB, BRN, and NIN backgrounds, respectively. Each nuisance is accompanied of its corresponding uncertainty, $\sigma$, which values are extracted from Ref.~\cite{DeRomeri:2022twg}.

\subsection{Argon}
Our second analysis corresponds to the LAr data provided by COHERENT~\cite{COHERENT:2020iec}. Here we compute the number of events in energy bins using Eq.~\eqref{eq:Nevents_CEvNS}, following the bin edges reported in~\cite{COHERENT:2020iec}. In this case, we extract the efficiency from the data release~\cite{COHERENT:2020ybo} and the smearing is modeled as a Gaussian function with an energy-dependent resolution as given in the same reference. Notice that the data in Ref.~\cite{COHERENT:2020iec} is given in units of keV$_{\rm ee}$. To properly use Eq.~\eqref{eq:Nevents_CEvNS} we convert between keV$_{\rm ee}$ and keV$_{\rm nr}$ by using the quenching factor reported in Ref.~\cite{COHERENT:2020iec} with a linear dependence on the nuclear recoil energy. For the LAr detector the collaboration has also released the data including timing information. Hence, we compute the expected number of events again in a two dimensional grid, each bin given by Eq.~\ref{eq:Nevents_CEvNS_ij}. However, there is no information regarding the time efficiency in this case and we take it equal to one. To perform our analysis, we minimize the squared function 
\begin{equation}
    \begin{aligned}
	\chi^2_{\mathrm{LAr}} = 
	\sum_{i,j}
	\frac{1}{\sigma_{ij}^2}\Bigl( 
     N_{ij}^\mathrm{th}(\mathfrak{n})  -  N_{ij}^{\text{exp}} \Bigr)^2  
     + \sum_{k=0,3,4,8}
	\left( 	\dfrac{ \gamma_{k} }{ \sigma_{\gamma_k} } 	\right)^2  +  \sum_{k=1,2,5,6,7} \gamma_{k} ^2 \,  ,
\end{aligned}
\label{chi2LAr}
\end{equation}
with $\sigma_{ij}^2 = N_{ij}^\text{exp} +  N_{ij}^\mathrm{SSB}/5$, where $N_{ij}^{\rm exp}$ are the measured number of events taken from the data release in Ref.~\cite{COHERENT:2020ybo} and $N_{ij}^{\rm th}$ are the expected number of events including backgrounds~\cite{DeRomeri:2022twg}
\begin{equation}
    \begin{aligned}
	N_{ij}^\mathrm{th}(\mathfrak{n}) =~
	 & (1 + \beta_0 +  \beta_1 \Delta_\mathrm{CE\nu NS}^{F_{90+}} + \beta_1 \Delta_\mathrm{CE\nu NS}^{F_{90-}} + \beta_2 \Delta_\mathrm{CE\nu NS}^\mathrm{t_{trig}}) N_{ij}^{S}  \\
    & + (1 + \beta_3) N_{ij}^\mathrm{SSB} + (1 + \beta_8) N_{ij}^\mathrm{dBRN}  \\
    & + (1 + \beta_4 + \beta_5 \Delta_\mathrm{pBRN}^{E_+} + \beta_5 \Delta_\mathrm{pBRN}^{E_-}
    + \beta_6 \Delta_\mathrm{pBRN}^{t_\text{trig}^+} + \beta_6 \Delta_\mathrm{pBRN}^{t_\text{trig}^-} + \beta_7 \Delta_\mathrm{pBRN}^{t_\text{trig}^\text{w}}) N_{ij}^\mathrm{pBRN}.\\
\end{aligned}
\label{chi2LAr}
\end{equation}
Notice that in this case the contribution from NIN backgrounds is neglected~\cite{COHERENT:2020iec} and the BRN component is separated in two contributions coming from prompt ($N_{ij}^{\rm pBRN}$) and delayed ($N_{ij}^{\rm dBRN}$) beam related neutrons. Following Ref.~\cite{DeRomeri:2022twg}, the nuiscane parameters, $\gamma_k$, are introduced to account for normalization ($k = 0, 3, 4, 8$) and shape uncertainties ($k = 1, 2, 5, 6, 7$), with the values of $\sigma_{\gamma_k}$ taken from the same reference. Coming back to Eq.~\eqref{chi2LAr}, the distributions $\Delta_\lambda^{\xi_\lambda}$ are defined as $ \Delta_{\lambda}^{\xi_\lambda} = (N_{ij}^{\lambda,\xi_\lambda} - N_{ij}^{\lambda,\mathrm{CV}})/{N_{ij}^{\lambda,\mathrm{CV}}}$, with CV the central value of each distribution~\cite{COHERENT:2020ybo, DeRomeri:2022twg}, and correspond to the $\pm1\sigma$ values of the corresponding distribution\footnote{More details on these distributions can be found in Ref.~\cite{DeRomeri:2022twg}}.
\subsection{Germanium}
For this measurement we consider the second data release from the COHERENT collaboration using Ge detectors \cite{COHERENT:2026yje}. Timing information for this experimental measurement has not been released yet. Hence, we compute the expected CEvNS events, $N_i^{\textrm{S}}$, by adding the three different flavour contributions, $N_{i,l}^{\textrm{S}}$ defined in Eq.~\eqref{eq:Nevents_CEvNS}, where we use a Gaussian smearing with a resolution as in Ref.~\cite{DeRomeri:2026dac}, an efficiency equal to one, and we convert between electron equivalent and nuclear recoil energy by adopting the Lindhard Model~\cite{osti_4701226, Lindhard:1961zz}. Once the expected events are computed, we perform the statistical analysis using the Poissonian $\chi^2$ function
\begin{equation}
    \chi^2=2\,\sum_i\left[N_i^{\textrm{th}} (\mathfrak{n}) - N_i^{exp}+N_i^{\textrm{exp}}\,\text{ln}\left(\frac{N_i^{\textrm{exp}}}{N_i^{\textrm{th}}(\mathfrak{n})}\right)\right]+\left(\frac{\eta}{\sigma_\eta}\right)^2+\left(\frac{\beta}{\sigma_\beta}\right)^2,
\label{eq:Chi_Ge}
\end{equation}
where
\begin{equation}
    N_i^{\textrm{th}}(\mathfrak{n})=(1+\eta)N_i^{\textrm{S}}(\mathfrak{n})+(1+\beta)N_i^{SSB},
\label{eq: Nth_Ge}
\end{equation}
with $N_i^{exp}$ the observed number of events in the $i$-th bin, and $N_i^{\textrm{SSB}}$ represents the steady state background taken from Ref.~\cite{COHERENT:2026yje}. The $\eta$ nuisance parameter is related to the systematic uncertainties on the signal, with $\sigma_\eta=10.3\%$~\cite{COHERENT:2026yje} and the parameter $\beta$ accounts for the SSB systematic uncertainty, with $\sigma_\beta=1\%$.

\begin{figure}[b]
    \centering
    \includegraphics[width=0.32\textwidth,height=0.22\textheight]{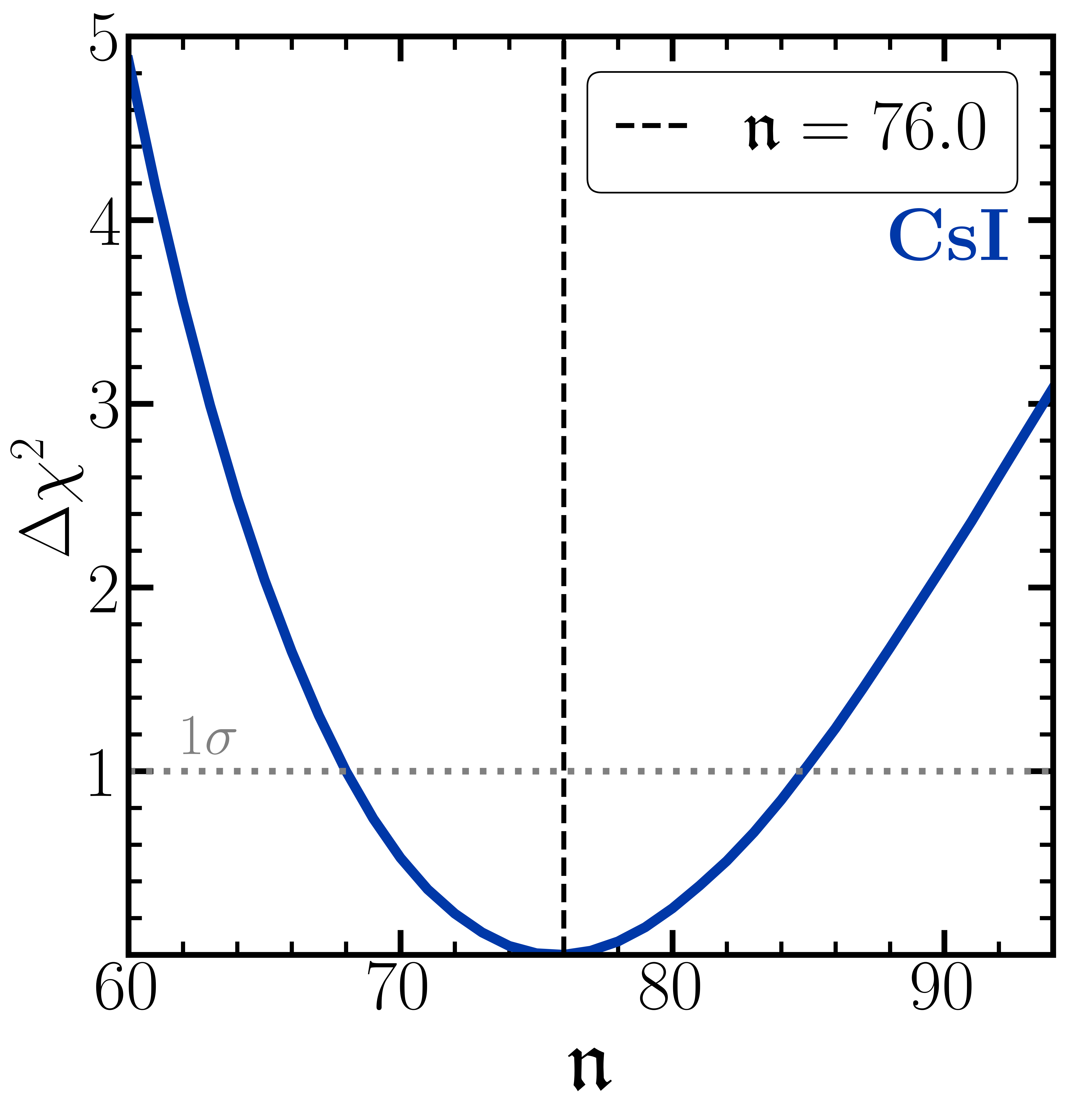}~~~
    \includegraphics[width=0.32\textwidth,height=0.22\textheight]{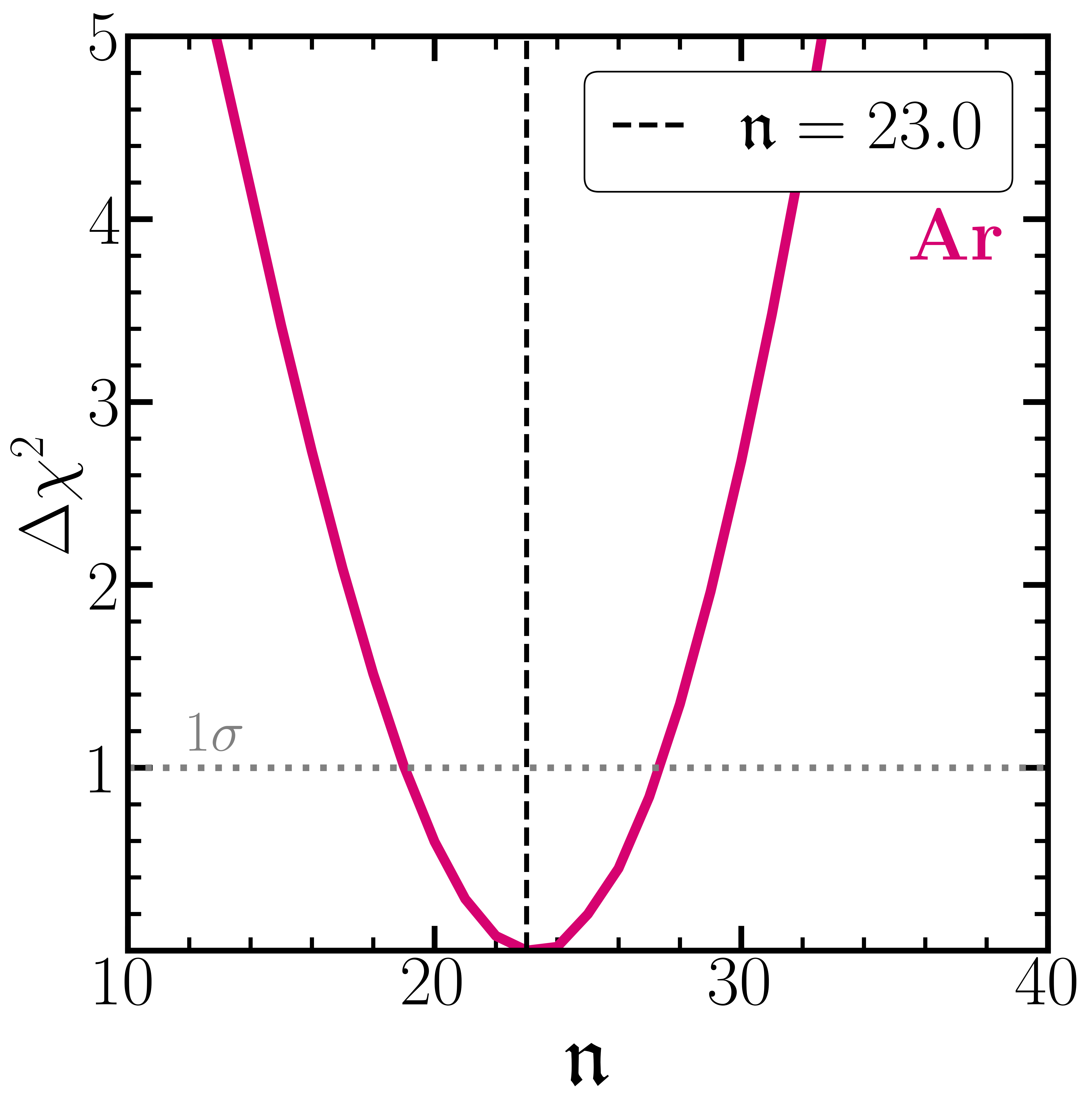}~~~
    \includegraphics[width=0.32\textwidth,height=0.22\textheight]{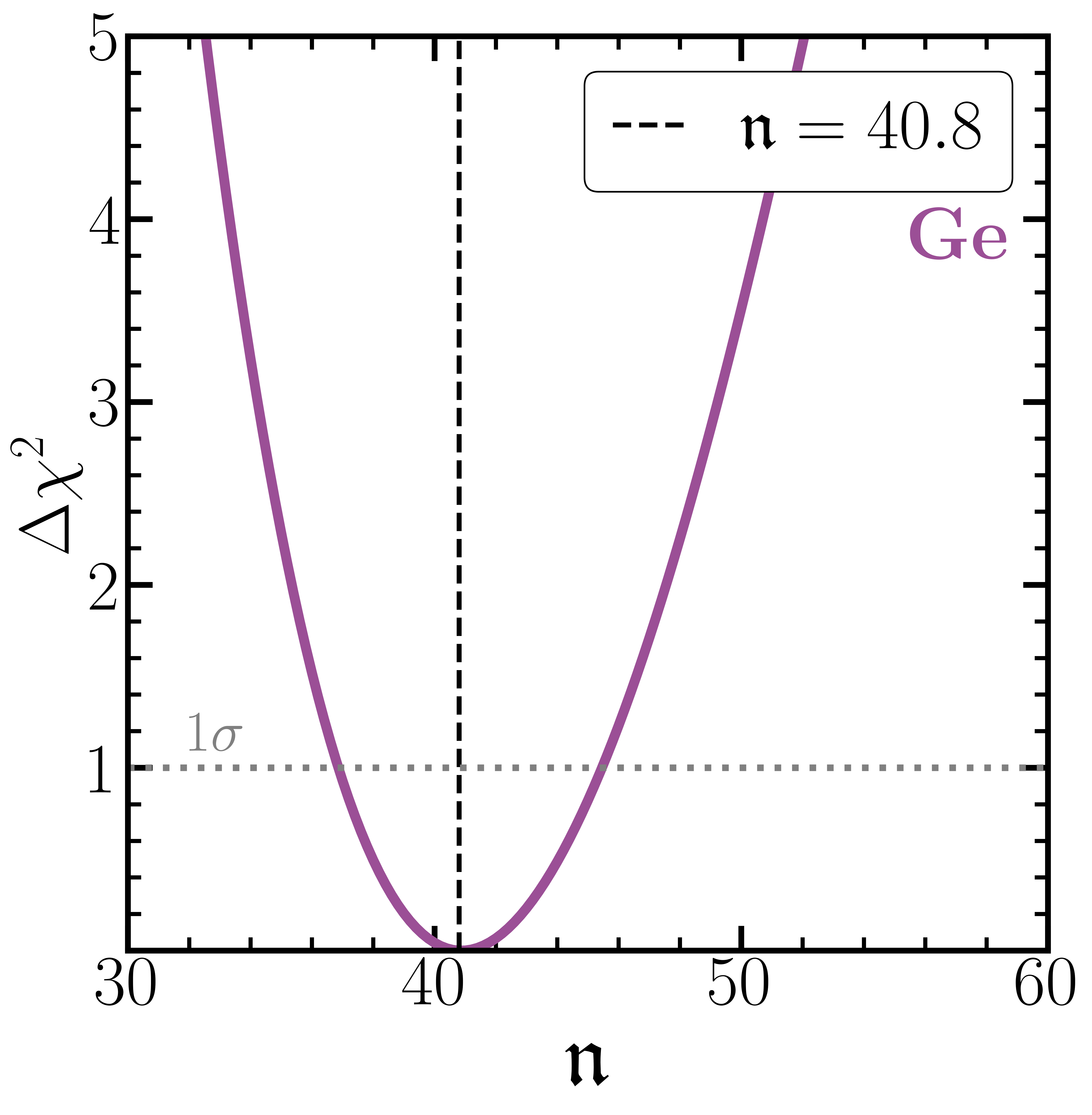}
    \caption{$\chi^2$ profiles for $N$ dependence for three datasets presented by the COHERENT collaboration. The best-fit values for CsI, Ar, and Ge targets are indicated by the vertical dashed lines at $\mathfrak{n}=76.0$, $\mathfrak{n}=23.0$, and $\mathfrak{n}=40.8$, respectively. The horizontal dotted line represents the $1\sigma$ confidence level.}
    \label{fig:Chi_profiles}
\end{figure}

\section{Results}

In this section we present the results of our analyses of the COHERENT data for the CsI, LAr, and Ge detectors. As discussed in section \ref{theory}, our main goal is to evaluate the dependence of the CEvNS cross section on the neutron number for each target material. We show in Fig.~\ref{fig:Chi_profiles} the $\Delta\chi^{2}$  profiles obtained by performing the analysis described in section~\ref{sec:analysis} for each experimental setup. The $1\sigma$ allowed intervals for each case are summarized in Table~\ref{tab:Results}, where we also show the average neutron number expected for each target~\footnote{For computing this average we consider that in CsI the proportion of Cs to I nuclei is 1:1, while for Ar we consider a single isotope-type, and for the case of Ge we compute an average weighted by the isotopic abundance of each Ge nuclei in natural germanium.}. From the table we see the extremely good agreement between the best fit and $N$.  These results show that, when considering the effective neutron number in the weak nuclear charge, nuclear form factors, and kinematic effects, the current CEvNS data are precise and accurate enough to reproduce the average number of neutrons in the target material using a single effective parameter.

\begin{figure}[h!]
    \centering
    \includegraphics[scale=0.3]{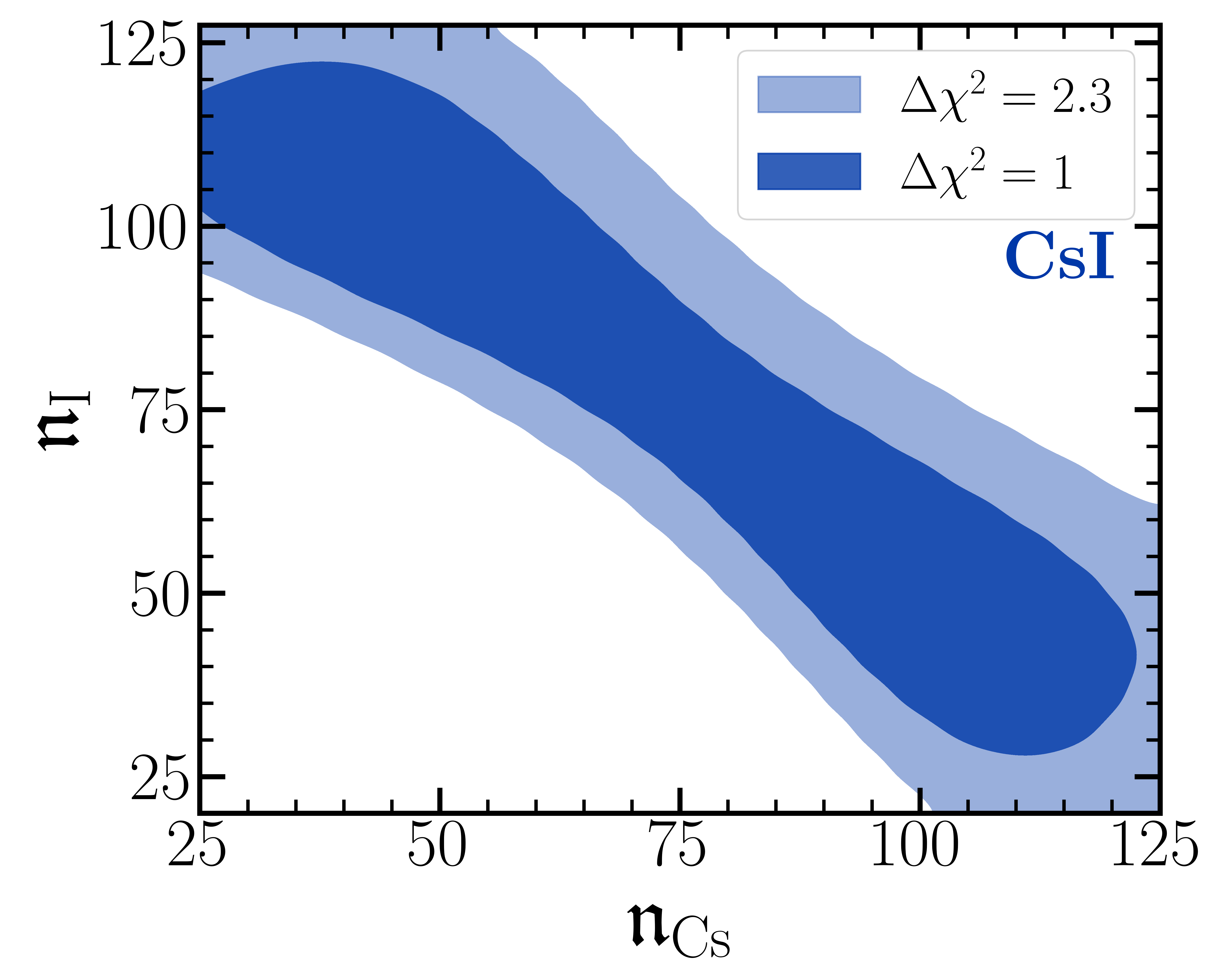}~~~\includegraphics[scale =0.3]{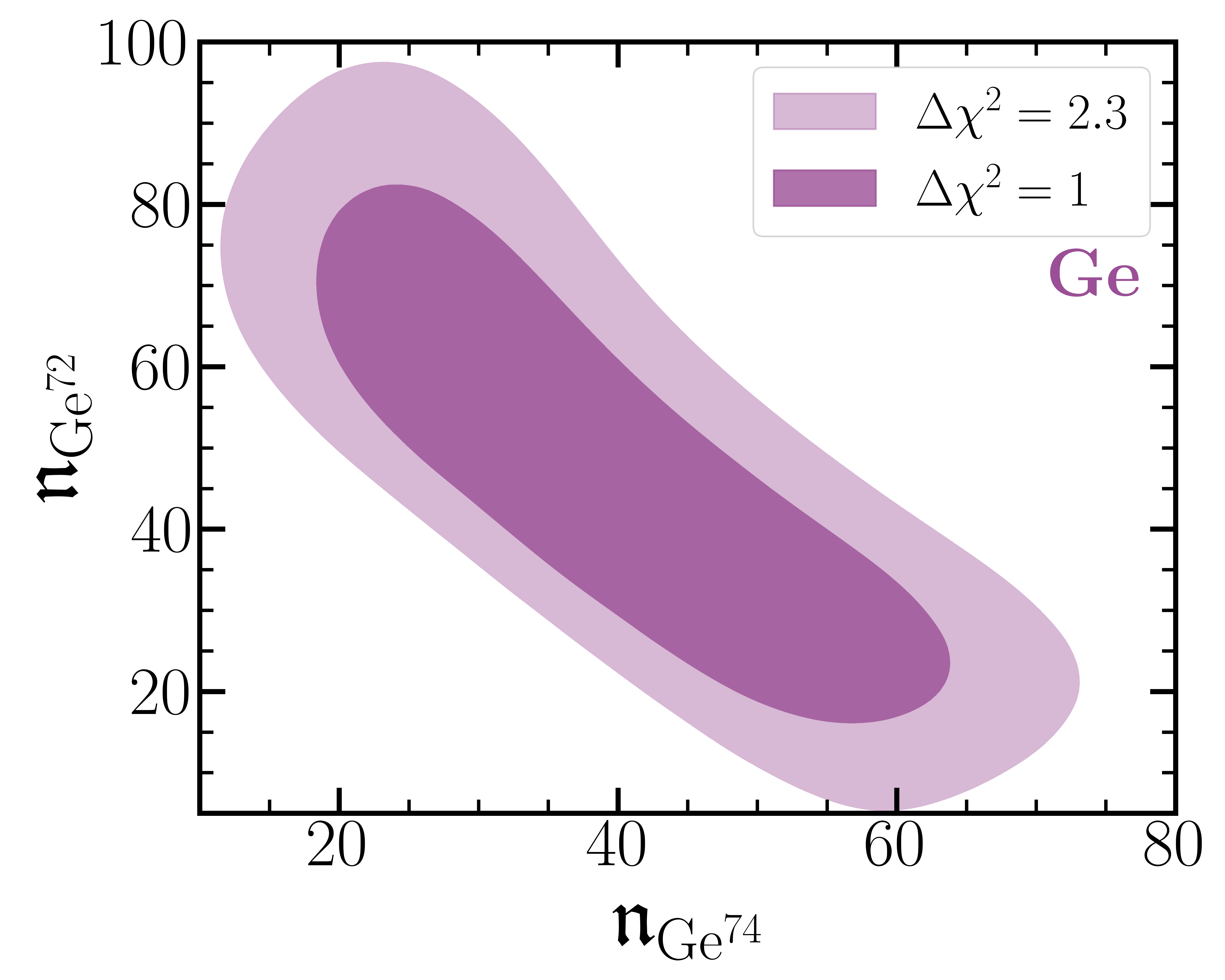}
    \caption{Two-dimensional confidence intervals for the effective neutron numbers of Cesium (Cs) and Iodine (I) (left panel) and for the Germanium isotopes $Ge^{74}$ and $Ge^{72}$ (right panel). The shaded regions denote the allowed parameter space at the $\Delta\chi^{2}=1$ (dark region) and $\Delta\chi^{2}=2.3$ (light region) significance levels.}
    \label{fig:Ge:2D}
\end{figure}

For completeness, we repeat our analysis now considering separate effective neutron numbers for the different constituent nuclei of the detector materials. The left panel in Fig.~\ref{fig:Ge:2D} shows the result for CsI, where we have defined an effective neutron number for Cs and I, independently. We can see that the point ($\mathfrak{n_{\textrm{Cs}}} = 78$, $\mathfrak{n_{\textrm{I}}} = 74$), corresponding to the actual neutron numbers for Cs and I, is consistent with our results at $1\sigma$ level, showing the accuracy of the experimental measurement. For the Ge detector, we recall that natural germanium consists of five stable isotopes. Hence, for the analysis we define an effective parameter only for those two with the largest relative abundance, i.e. $^{72}$Ge and $^{74}$Ge, and fix the other three to the actual value of $N$. The result is shown in the right panel of  Fig.~\ref{fig:Ge:2D}, where we see that the point ($\mathfrak{n_{\textrm{Ge}^{\textrm{74}}}} = 42$, $\mathfrak{n_{\textrm{Ge}^{\textrm{72}}}} = 40$), the actual neutron numbers, is again in agreement at a $1\sigma$ level.

\begin{table}[t]
\centering
\begingroup
\renewcommand{\arraystretch}{1.4}
\begin{tabular}{|c|c|c|c|c|c|}
\hline
 & ~~~Effective Neutron Best Fit ($\mathfrak{n}$)~~~ & ~~~Average Neutron Number ($N$)~~~  \\ \hline
\textbf{CsI} & $76.0 ^{+8.82}_{-8}$ & 76   \\ \hline
\textbf{Ar} & $23.0^{+4.32}_{-3.98}$ & 22  \\ \hline
\textbf{Ge} & $40.8^{+4.66}_{-3.94}$ & 40.7  \\ \hline
\end{tabular}
\endgroup
\caption{Results for the effective neutron number ($\mathfrak{n}$), uncertainties correspond to 1$\sigma$ confidence level. We also show, for comparison, the average neutron number for each material. }
\label{tab:Results}
\end{table}

\section{Conclusions}

With the results shown here, we demonstrate that current SNS experiments, particularly the latest germanium data, confirm the $N^2$-rule for the CEvNS cross section. The accuracy and precision of these type of experiments is illustrated by measuring the effective average number of neutrons in the target detector with the use of the $N$-dependence of the CEvNS cross section. This particular dependence within the CEvNS reaction provides a particular accuracy test that can be used as a benchmark for the precision of the CEvNS experiments, a tool that is not available in other experiments.

\section*{Acknowledgments}
\noindent O. G. M. and G. S. G. would like to thank SNII (Sistema Nacional de Investigadoras e Investigadores, Mexico. G.S.G. has been supported by SECIHTI Project No. CBF-2025-I-1589.

%\bibliographystyle{utphys}
%\bibliography{bibliography}  

\providecommand{\href}[2]{#2}\begingroup\raggedright\endgroup

\end{document}